\documentclass[aps,prl,showpacs,twocolumn,superscriptaddress,a4paper]{revtex4}
\usepackage[pdftex]{graphicx}
\usepackage{epstopdf}
\usepackage{amsmath}
\usepackage{epsfig}
\usepackage{float}

\usepackage[T1]{fontenc}
\usepackage[latin9]{inputenc}
\usepackage{subfigure}
\usepackage{textcomp}

\providecommand{\tabularnewline}{\\}

\begin{document}

\title{First-principles study of ferroelectric domain walls in multiferroic
bismuth ferrite}

\date{\today}

\author{Axel Lubk}
\affiliation{Institute for Structure Physics, Technische Universitaet Dresden, D-01062, Germany}
\email{axel.rother@triebenberg.de}

\author{S.\ Gemming}
\affiliation{Institute of Ion-Beam Physics and Materials Research, FZ Dresden-Rossendorf, D-01328 Dresden}
\email{s.gemming@fzd.de}

\author{N.~A.\ Spaldin}
\affiliation{Materials Department, University of California, Santa Barbara, CA,
93106-5050, USA}
\email{nicola@mrl.ucsb.edu}

\begin{abstract}
We present a first-principles density functional study of the structural,
electronic and magnetic properties of the ferroelectric domain walls in 
multiferroic BiFeO$_3$. We find that domain walls in which the rotations of 
the oxygen octahedra do not change their phase when the polarization 
reorients are the most favorable, and of these the 109$^{\circ}$ 
domain wall centered around the BiO plane has the lowest energy.
The 109$^{\circ}$ and 180$^{\circ}$ walls have a significant change in
the component of their polarization perpendicular to the wall; the 
corresponding step in the electrostatic potential is consistent with
a recent report of electrical conductivity at the domain walls. Finally,
we show that changes in the Fe-O-Fe bond angles at the domain walls
cause changes in the canting of the Fe magnetic moments which
can enhance the local magnetization at the domain walls.
\end{abstract}


\maketitle

\section{Introduction}

Perovskite-structured bismuth ferrite, BiFeO$_{3}$, is the subject
of much current research because of its large room temperature ferroelectric
polarization and simultaneous (antiferro-)magnetic ordering. Such
\textit{multiferroic} materials show a wealth of complex physical
properties caused by their coexisting electrical and magnetic order
parameters, which in turn suggest potential applications in novel
magnetoelectronic devices: Recent reports of electric-field induced
switching of magnetization through exchange bias of ferromagnetic
Co to BiFeO$_{3}$ are particularly promising \cite{Zhao_et_al:2006,Chu_et_al:2008}.
In addition, the large ferroelectric polarization \cite{Wang_et_al:2003},
driven by the stereochemically active Bi$^{3+}$ lone pair \cite{Neaton_et_al:2005},
is motivating investigation of its purely ferroelectric behavior for
possible applications in ferroelectric memories.

The suitability of ferroelectric materials for applications
is determined not only by the magnitude of their ferroelectric polarization,
but also by factors such as switchability, fatigue and loss. These
are in turn influenced by the structure of the domains and particularly
by the boundaries between them. The detailed structure and formation
energies of domain walls in some conventional ferroelectrics are now
well established (see for example Refs. \onlinecite{Stemmer(1995),Floquet/Valot:1999}
for experimental studies and Refs. 
\onlinecite{Padilla/Zhong/Vanderbilt:1996,Meyer/Vanderbilt:2002}
for calculations). For BiFeO$_3$, however, the first experimental study of 
domain walls was only recently reported \cite{Seidel(2009)}, and a detailed 
theoretical study is lacking. The additional consideration of the effect of the 
ferroelectric domain wall on the electronic and magnetic degrees of freedom makes such a study 
particularly compelling.

In this work we use density functional theory within the LSDA$+U$
method to calculate the structure, stability and properties of the
ferroelectric domain walls in BiFeO$_{3}$. We begin this manuscript by reviewing the 
structure of bulk BiFeO$_3$ (Section~\ref{Structure-model}), so that we can
use the bulk symmetry to determine the allowed energetically favorable
domain wall orientations 
(Section~\ref{Symmetry}). In Section~\ref{DFT-calculations} we describe
the technical details of our density functional calculations. The main
part of the paper -- Section~\ref{Results} -- contains our results:
We perform full structural optimizations of the atomic positions for the 
energetically favorable domain wall orientations, and calculate and
compare their total energies to determine which walls are most likely
to occur.
We then calculate the electronic and magnetic properties of the walls, 
paying particular attention to how changes in structure at the boundaries
influence the electronic densities of states, potential profile and spin
canting. The implications of our findings for domain walls in multiferroics,
and suggestions for future directions are summarized in Section~\ref{Summary}.

\section{Structure of B\lowercase{i}F\lowercase{e}O$_{3}$\label{Structure-model}}

\begin{figure}
\includegraphics[scale=1]{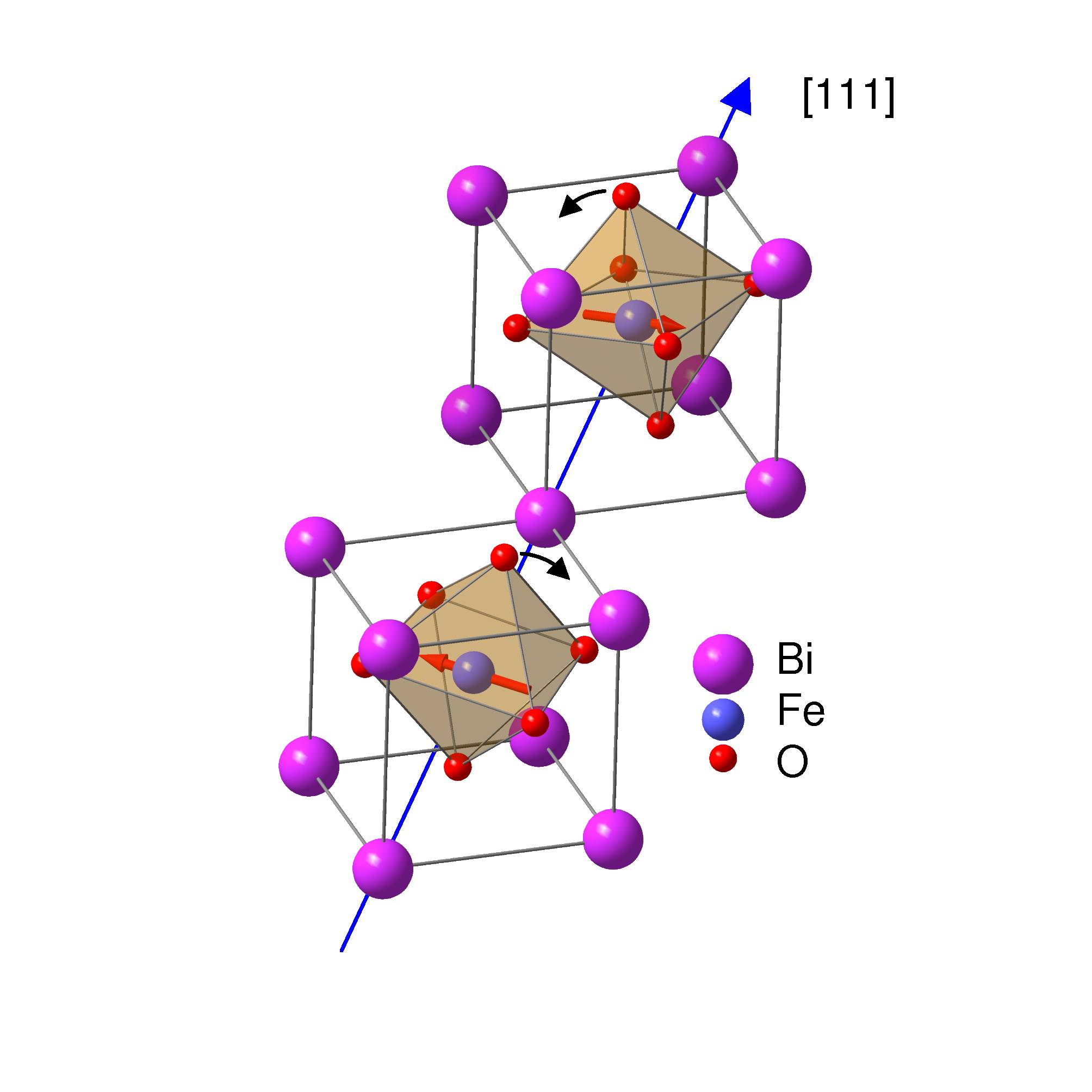}
\caption{Crystal structure of bulk BiFeO$_{3}$. Two simple perovskite
unit cells are shown to illustrate that successive oxygen octahedra 
along the polar $[111]$ axis rotate with opposite sense around $[111]$. 
The red arrows on the Fe atoms indicate the orientation of the 
magnetic moments in the $(111)$ plane.
\label{fig:unit cell}}
\end{figure}

BiFeO$_{3}$ is a rhombohedral perovskite with space group $R3c$ (Fig. \ref{fig:unit cell}).
The ground state structure is reached from the ideal cubic perovskite ($Pm\bar{3}m$)
by imposing two symmetry-adapted lattice modes: (1) a non-polar $R$-point
mode which rotates successive oxygen octahedra in opposite sense
around the $[111]$-direction, and (2) a polar $\Gamma_{4}^{-}$ distortion,
consisting of polar displacements along the $[111]$-direction as
well as symmetric breathing of adjacent oxygen triangles \cite{Fennie_2:2008}.
The rhombohedral lattice constant is 5.63 \AA\ (with 
corresponding pseudocubic lattice constant, $a_0 = 3.89$ \AA), and the 
rhombohedral angle, $\alpha=59.35^{\circ}$ is close to the 
ideal value of $60^{\circ}$ \cite{Kubel/Schmid:1990}.
First-principles density functional calculations have been shown to
accurately reproduce these values \cite{Neaton_et_al:2005}.

The magnetic ordering is well established to be G-type antiferromagnetic
\cite{Fischer_et_al:1980}, with a long wavelength spiral of the
antiferromagnetic axis \cite{Sosnowska/Peterlin-Neumaier/Streichele:1982}.
First-principles density functional computations \cite{Ederer/Spaldin:2005}
and symmetry considerations \cite{Fennie:2008} indicate a local
canting of the magnetic moments to yield a weak ferromagnetic moment;
this canting is symmetry allowed because of the presence of the non-polar 
$R$ point rotations of the oxygen octahedra. Since the orientation of the 
weak ferromagnetic moment follows the axis of the long-range spiral no net 
magnetization results.

The ferroelectric polarization is large, $\sim$90$\mu$C/cm$^{2}$
\cite{Wang_et_al:2003}, and can point along any of the eight pseudo-cubic
$\left\langle 111\right\rangle $ directions \cite{Kubel/Schmid:1990}.
Simple geometrical considerations therefore suggest angles of $\pm71^{\circ}$,
$\pm109^{\circ}$ or $180^{\circ}$ between allowed polarization orientations 
of the ideal rhombohedral system ($\alpha=60^{\circ}$) \cite{Streiffer1998}; 
we will label our domain walls using these angles in the following.  
Experimentally, such relative domain orientations and re-orientation
angles have indeed been observed \cite{Zhao_et_al:2006}. Currently
nothing is known about the behavior of the octahedral rotations or
the magnetism at the domain boundaries.

\section{Symmetry analysis of domain walls\label{Symmetry}}

In general, the energetically favorable domain wall configurations 
for a particular symmetry are those orientations which can be free
of both stress and space charge. For the rhombohedral symmetry of
BiFeO$_3$, these conditions lead to the following likely domain wall 
orientations for $\pm71^{\circ}$, $\pm109^{\circ}$ and $180^{\circ}$
orientations respectively: 
$\left\{ 011\right\}$, $\left\{ 001\right\}$ and $\left\{ 011\right\}$ 
(in pseudo-cubic coordinates) \cite{Streiffer1998}.
For each of these wall orientations, there is a choice of atomic plane
about which the initial domain wall can be centered (for example around a BiO
or FeO$_2$ plane in the 109$^{\circ}$ case). In addition, since the rotations 
of the oxygen octahedra
are uncoupled from the orientation of the polarization, different relative
orientations of oxygen octahedra on either side of the domain wall are 
possible. In order to survey all possibilities we investigate the 
following domain boundaries:

\begin{enumerate}
\item $71^{\circ}$: We construct the domain wall in the $(011)$ plane 
with the electric polarization changing from the $[111]$ direction on one 
side of the domain wall to $[\bar{1}11]$ on the other (Fig. \ref{fig:struc}(a)).
We study two configurations of the rotations of the oxygen octahedra, which
we refer to as either {\it continuous} or {\it changing}. In the continuous 
case, the phase of the oxygen octahedral rotations remains unchanged along an
integral curve of the polarization vector field; in the changing case the
phase reverses at the domain wall. In principle the wall could be centered
around either a BiFeO or O$_{2}$ plane (or any intermediate plane). However, 
since the domain wall location is not fixed by symmetry and the distance 
between the two planes is small, the fully relaxed domain boundary will likely
be centered close to the O$_{2}$ plane independent of the initial configuration 
(as found in Ref. \cite{Meyer/Vanderbilt:2002}).

\begin{figure}
\subfigure[]{\includegraphics[scale=0.5]{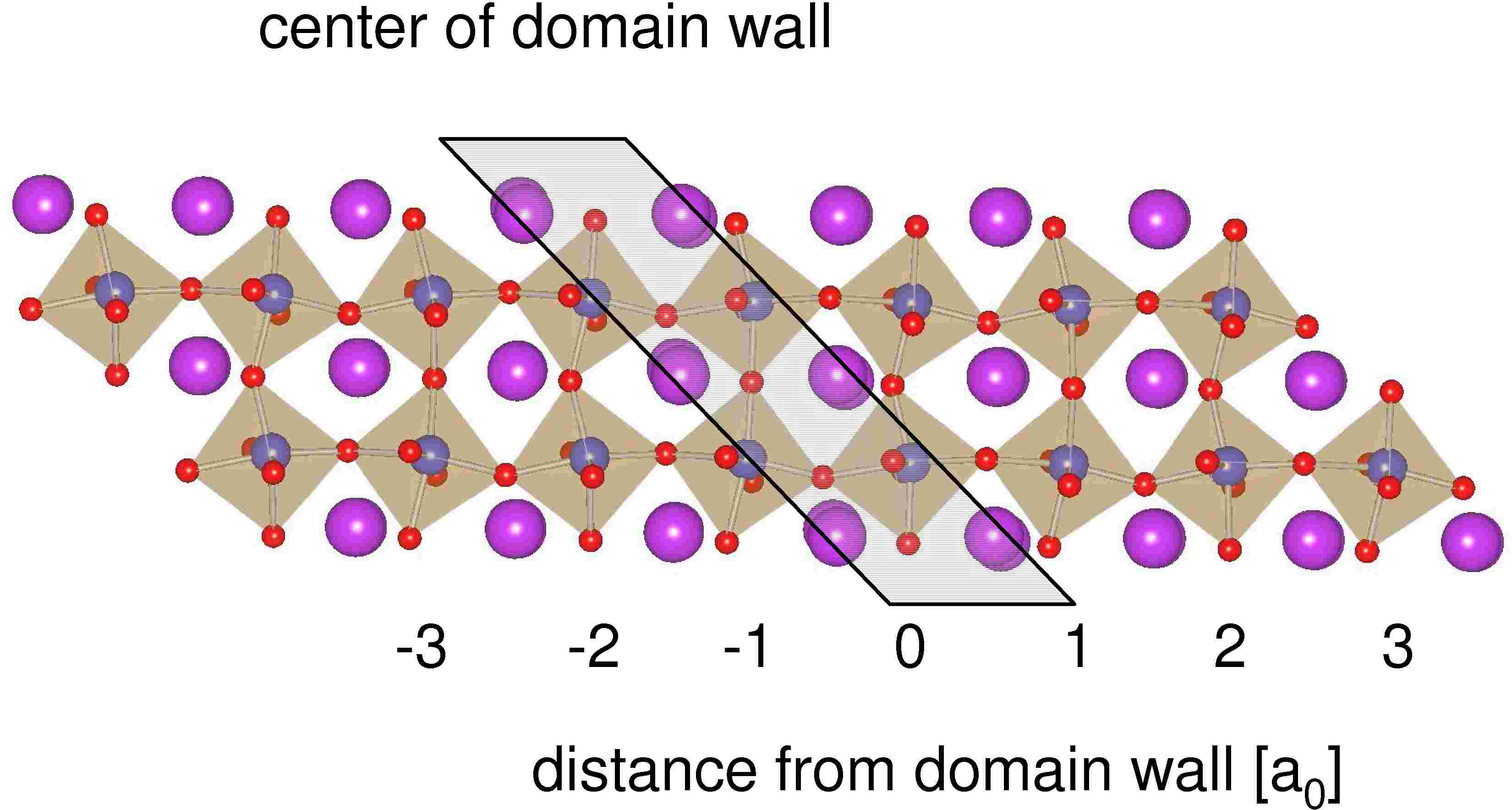}}
\subfigure[]{\includegraphics[scale=0.5]{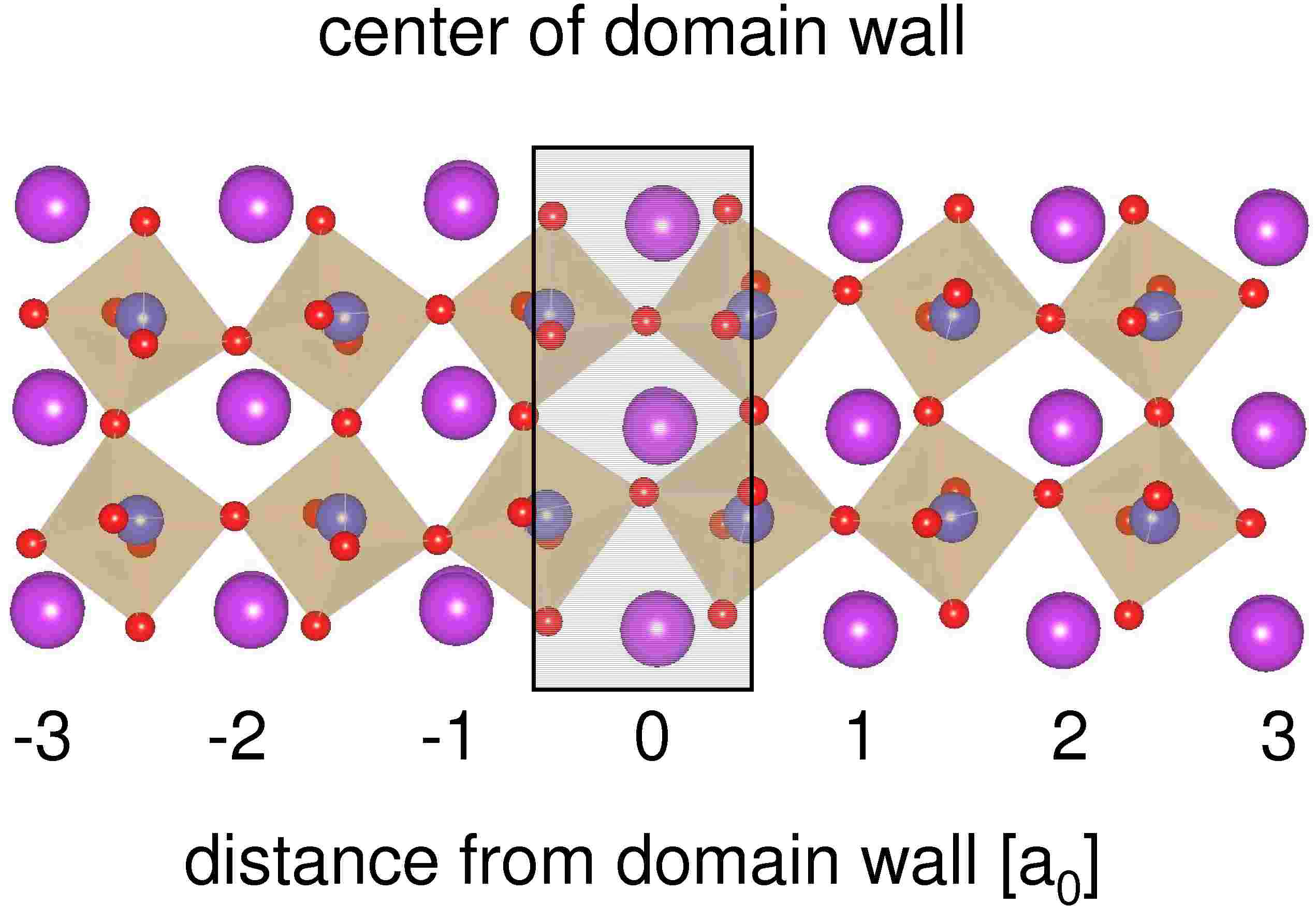}}
\subfigure[]{\includegraphics[scale=0.5]{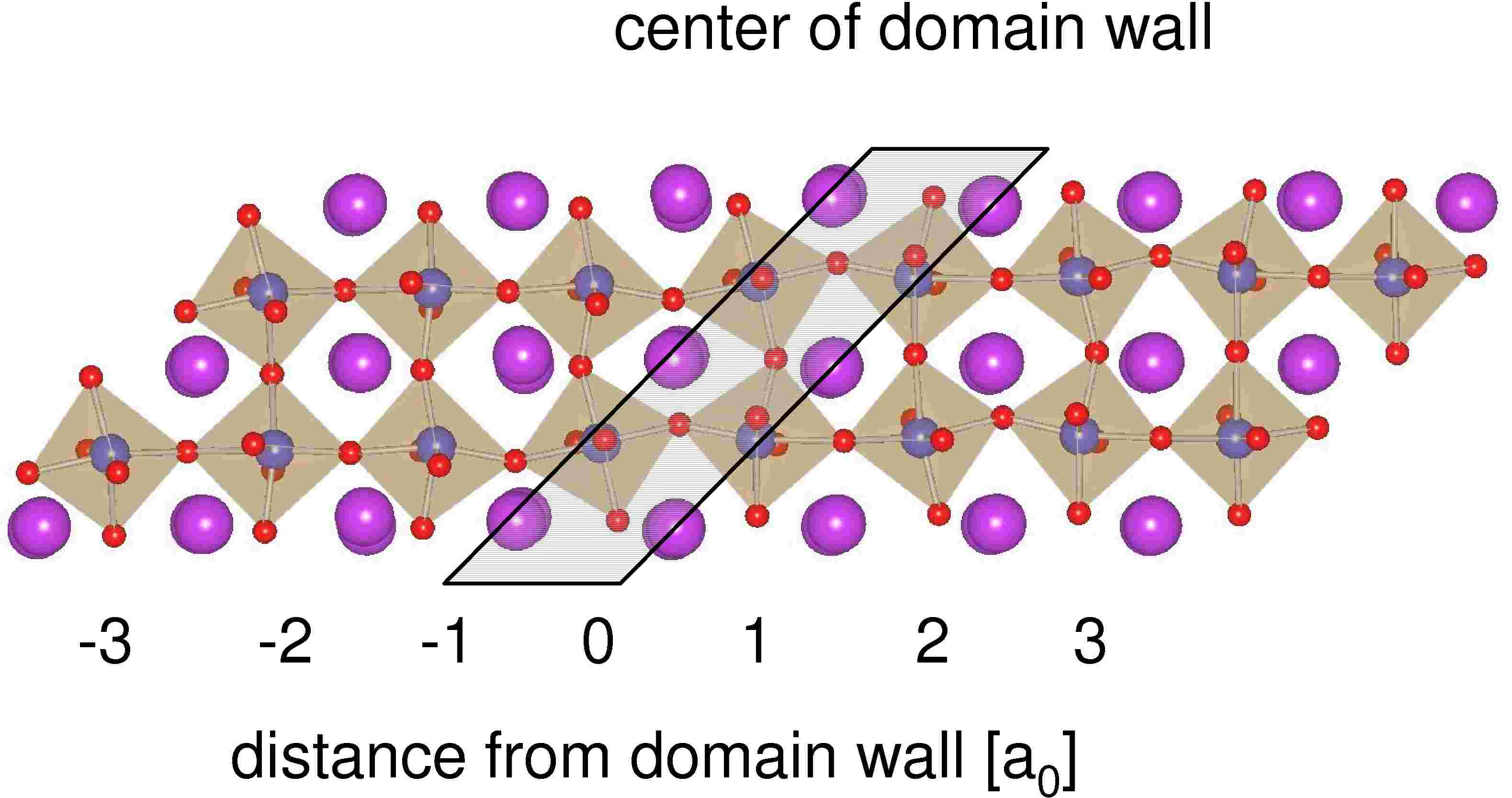}}
\caption{(a) $71^{\circ}$ domain boundary with continuous oxygen octahedral
rotations. (b) $109^{\circ}$ domain boundary with continuous oxygen octahedral
rotations centered on the FeO$_2$ plane. (c) 180\textdegree{} domain boundary with continuous oxygen octahedral
rotations. Note that only half of the supercell is shown. 
\label{fig:struc}}
\end{figure}

\item $109^{\circ}$: We use the $\left(001\right)$ plane for the domain wall, with
polarization changing from the $[111]$ to the $[\bar{1}\bar{1}1]$ direction
(Fig. \ref{fig:struc}(b)). Again we explore two configurations of the octahedral 
rotations, with continuous or changing phase along the integral curve of the polarization vector field. 
In this case we were able to separately resolve domain walls centered
on FeO$_{2}$ and BiO planes, since their separation is $\sqrt{2}$ times
that of the BiFeO and O$_{2}$ planes in the 71$^{\circ}$ case.

\item $180^{\circ}$: We use the $(0\bar{1}1)$ plane for the domain wall, with polarization 
changing between $[111]$ and $[\bar{1}\bar{1}\bar{1}]$ directions (Fig. \ref{fig:struc}(c)). 
Again we explore
two octahedral tilt patterns, with the phases along the polarization direction
either continuous or changing across the boundary. 
As in the $71^{\circ}$ case, we 
do not distinguish between the BiFeO- and O$_{2}$-centered domain walls.

\end{enumerate}

\section{Computational details\label{DFT-calculations}}

We performed density functional (DFT) calculations using the Vienna 
\it ab-initio \rm simulation package, VASP \cite{Kresse/Furthmueller_CMS:1996}.
We used the projector augmented wave method \cite{Bloechl:1994,Kresse/Joubert:1999} 
with the default VASP PAW potentials including semi-core states in the 
valence manifold (core states Bi: [Kr], Fe: [Ne]$3s^2$, O: $1s^2$). 
We used the rotationally invariant implementation \cite{Liechenstein/Anisimov/Zaanen:1995}
of the LSDA+U method \cite{Anisimov/Aryasetiawan/Liechtenstein:1997} to describe the 
exchange-correlation functional with values of $U=3$ eV and $J=1$ eV
that were shown previously to accurately reproduce the experimentally
observed structural and electronic properties of bulk BiFeO$_3$ 
\cite{Neaton_et_al:2005,Ederer/Spaldin:2005,Ederer/Spaldin_2:2005}.

We constructed supercells containing two domains separated by domain walls,
with a total of 120 atoms (60 atoms per domain); the width of each domain 
was then six pseudocubic unit cells. We used the lattice parameters obtained
from calculations for bulk BiFeO$_3$ (see Ref. \onlinecite{Neaton_et_al:2005}), 
with a slight change: the rhombohedral angle, $\alpha$, was taken to be exactly
$60^{\circ}$ to allow us to incorporate both domains in one supercell. 
While the electronic structure of bulk BiFeO$_3$ at $\alpha = 60^{\circ}$
is indistinguishable from that at the experimental $\alpha = 59.35^{\circ}$
\cite{Neaton_et_al:2005},
we point out that this constraint might influence the strain profile at the
domain boundary. The total energy difference of bulk BiFeO$_3$ with $\alpha = 60^{\circ}$ and with $\alpha = 59.35^{\circ}$ is below 1 meV, hence we conclude, that the effect may be neglected in the further discussion.
We initialized the magnetic ordering to the $G$-type antiferromagnetic 
arrangement known to occur in the bulk. 

Full structural optimizations of the atomic positions (until the forces on
each ion were below 0.03 eV per \AA) and cell parameters (until energy differences
were below 0.01 eV) were then performed for all of the domain configurations 
described in Section~\ref{Symmetry}. No symmetry constraints were imposed. 
The cell parameter relaxations were necessary because the interlayer distance
in all three domain walls is slightly larger (by around 0.1 \AA) than that
in the bulk. The rather large remaining forces occur due to a combination of the complicated
 crystal structure of BiFeO$_3$ and the large number of atoms in one supercell, leading to 
a flat energy surface and a particularly slow structural convergence. Additionally, special care had to be taken with respect to the starting conditions of the structure 
optimization, i.e. several different initializations of the initial spin configuration and ion 
positions were performed for each configuration to reduce the probability of 
being trapped in local minima. 
We used 5$\times$3$\times$1 (71$^{\circ}$ and 180$^{\circ}$) 
and 5$\times$5$\times$1 (109$^{\circ}$) {\it k}-point samplings; these correspond
to values that have been shown to give good convergence for bulk BiFeO$_3$,
with a denser sampling along the long axis of the supercell.
The plane wave energy cut-off was set to 550 eV. 

Finally, for the calculated lowest energy $109^{\circ}$ boundary, we performed 
additional non-collinear magnetic calculations with spin-orbit coupling included.

\section{Results\label{Results}}

\subsection{Structure and energetics}

In all cases our supercells relaxed to contain two distinct domains,
with the layers in the middle of each domain having similar structure
to that of bulk BiFeO$_3$; this suggests that the supercells were
large enough to minimize interactions between the domain walls.

\begin{table}
\begin{tabular}{|c|c|c|c|c|c|c|}
\hline 
$71^{\circ}$ c & $71^{\circ}$ d & $109^{\circ}$ Bc & $109^{\circ}$ Bd & $109^{\circ}$ Fc & 
$109^{\circ}$ Fd & $180^{\circ}$ c\tabularnewline
\hline
\hline 
363 & 436 & 205  & 896  & 492  & 1811  & 829\tabularnewline
\hline
\end{tabular}
\caption{Calculated domain wall energies (mJ/m$^{2}$) for 71$^{\circ}$, 
109$^{\circ}$ and 180$^{\circ}$ walls.  B and  F indicate the BiFeO- and 
FeO$_2$-centered planes, c and d label the continuous or discontinuous oxygen 
octahedral rotations.}
\label{tab:domain-wall-energies}
\end{table}
In Table~\ref{tab:domain-wall-energies} 
we list our calculated domain wall energies for
all of the configurations described in the previous section. It
is clear that, in all cases, the configuration with the least
perturbation to the phase of the octahedral rotations is lowest
in energy. Indeed, in the 180$^{\circ}$ case we were unable to
obtain a converged solution for the case with reversal of the 
octahedral rotations at the domain boundary. The large differences between 
the continuous and discontinuous oxygen octahedra rotations is a peculiarity 
of the BiFeO$_{3}$ structure and indicates the importance of the Fe-O-Fe 
bonding angles in determining the structural stability. The 109$^{\circ}$  
wall is energetically the most stable of the three orientations.
It is somewhat surprising that the 109$^{\circ}$ wall is lower in energy
that the 71$^{\circ}$ wall; since the change in orientation of the electric
polarization vector is smaller in the latter, one would also expect the 
perturbation to the structure to be smaller. (Previous calculations for 
PbTiO$_3$ found the 90$^{\circ}$ wall to be lower in energy than the 
180$^{\circ}$ wall, consistent with this argument \cite{Meyer/Vanderbilt:2002}).
We believe that this reversal is caused by the favorable arrangement of the oxygen 
octahedra at the 109$^{\circ}$ wall boudary: 
since the 109$^{\circ}$ wall lies in
the $\left\{ 001\right\}$ plane, it is oriented along the apices
of the oxygen octahedra (Fig.~\ref{fig:struc}(b) upper panel),  
whereas the 71$^{\circ}$ and 180$^{\circ}$ $\left\{ 011\right\}$ 
walls are oriented along the octahedral edges (Figs.~\ref{fig:struc}(a) 
and ~\ref{fig:struc}(c) upper panels) giving them less freedom 
to accommodate the changes in polarization direction. 
The 109\textdegree{} BiO-centered wall is lower in energy than 
the FeO$_{2}$-centered wall, consistent with previous studies 
for other perovskite ferroelectric domain boundaries which also found
AO-centered walls to be more stable \cite{Meyer/Vanderbilt:2002}.
The 180$^{\circ}$ case has the highest domain wall energy, consistent
with its having the largest change in the polarization orientation.   
Finally, we note that the domain wall energies in BiFeO$_3$ are 
significantly larger than those calculated for PbTiO$_3$, which
in turn are larger than the BaTiO$_3$ values (Table 
~\ref{tab:domain-wall-energies2}). The large increase from BaTiO$_3$
to PbTiO$_3$ suggests a correlation between polarization magnitude
and domain wall energy. While changes in polarization would predict 
somewhat larger domain wall energies for BiFeO$_3$, there is a large 
additional increase which is likely a result of the additional deformations
caused by the octahedral rotations (see discussion above). It is
also possible that the magnetic energy cost associated with perturbing
the Fe-O-Fe bond angles further raises the domain wall energies in BiFeO$_3$. 
\begin{table}
	\centering
		\begin{tabular}{|c|c|c|}
		\hline 
		angle  & $BaTiO_3$ & $PbTiO_3$ \tabularnewline
		\hline
		\hline 
		90° & N/A & 35.2\tabularnewline
		\hline
		180° & 7.5 & 132\tabularnewline
		\hline
	\end{tabular}
	\caption{Lowest calculated domain wall energies (mJ/m$^{2}$) for 90° and 180° domain walls in BaTiO$_{3}$ and PbTiO$_{3}$, from Ref. \onlinecite{Meyer/Vanderbilt:2002}. }
	\label{tab:domain-wall-energies2}
\end{table}

\begin{figure}
\includegraphics[scale=0.8]{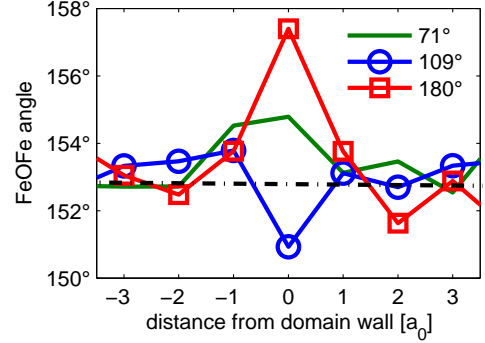}
\caption{Fe-O-Fe angles in each layer of the supercell. The bulk value of 152.9$^{\circ}$
is indicated by the dashed line.  Note the changes in angle in the domain wall region.
\label{fig:Fe-Fe-distances-across}}
\end{figure}

As a measure of the amount of structural distortion,
in Figure \ref{fig:Fe-Fe-distances-across} we plot the Fe-O-Fe angles
in each layer across the supercells. Within the central region of the domain 
the bulk value of 152.9$^{\circ}$ is regained as expected. Indeed
the bulk behavior is recovered
within one or two layers of the domain wall boundary, consistent 
with earlier studies on PbTiO$_{3}$ domain walls [\onlinecite{Meyer/Vanderbilt:2002}]. 
The angles change by up to $\sim$4$^{\circ}$ in the wall region to accommodate
the changes in structure associated with the polarization reorientation. 
However, the Fe-O-Fe angles remain far from 180$^{\circ}$ in all cases,
indicating that the structure within the walls is far from an ideal cubic 
perovskite structure. 
Since the Fe-O-Fe angle strongly influences the superexchange interactions
and the local anisotropy,
we anticipate that these changes in angles might influence the magnetic 
properties; we return to this point later. 

\subsection{Evolution of the polarization across the domain walls}

In order to better understand the change in structure across the 
domain wall we performed a layer-by-layer analysis of the local 
polarization by summing over the displacements of the atoms in each
layer from their ideal cubic perovskite positions, multiplied by
their Born effective charges (BECs). While there is not a unique
way to partition the layers, we find that our results from different 
decomposition schemes are similar, and so we use the narrowest possible
layer partition in order to optimize the resolution.
We used the BECs of the $R3c$ structure calculated
in a previous study \cite{Neaton_et_al:2005} using the same computational
parameters as we use here; note that the actual BECs might deviate
slightly from these values. This technique was used previously
to analyze the polarization evolution across PbTiO$_3$ domain walls
\cite{Meyer/Vanderbilt:2002}. 
We are particularly interested in
two factors: First whether the polarization reorientation takes
place through a rigid rotation of the local polarization, without
a reduction in its local magnitude (analogous to the rotation of
a magnetic moment in a Bloch wall in a ferromagnet). And second, 
whether a change in polarization in the direction perpendicular
to the wall develops. This is of particular interest since, as discussed
in earlier work \cite{Meyer/Vanderbilt:2002}, it gives rise to a potential step
at the boundary which, if screened by a dipole layer in the charge 
density, could give rise to intriguing effects such as enhanced 
conductivity at the boundary.

\begin{figure}
\includegraphics[scale=0.8]{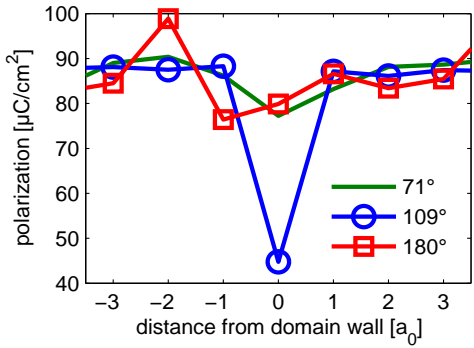}
\caption{Layer-by-layer polarization calculated from the sum of the displacements
of the ions from their ideal positions multiplied by the Born effective charges.}
\label{fig:pol}
\end{figure}

First, in Fig.~\ref{fig:pol} we show the {\it magnitudes} of the calculated 
layer-by-layer polarizations for all three wall types. 
It is clear that in the 71$^{\circ}$ and 180$^{\circ}$ walls, the magnitude of
the polarization remains approximately constant across the wall, indicating
a rigid rotation of the polarization in the manner of a magnetic Bloch wall.
(Note that the scatter in the local polarization especially at the 180$^{\circ}$
results from our fairly high force tolerance of 0.03 eV per \AA.)
In contrast, the 109$^{\circ}$ wall has a marked reduction in the 
local polarization in the wall region; this likely results from the 
greater structural flexibility provided by the orientation of the 
109$^{\circ}$ wall relative to the corners of the octahedra. 

\begin{figure}
\subfigure[]{\includegraphics[scale=0.8]{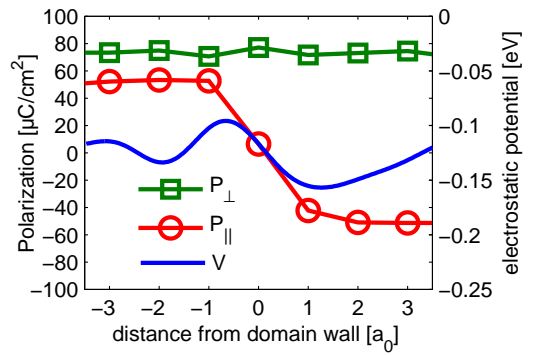}}
\subfigure[]{\includegraphics[scale=0.8]{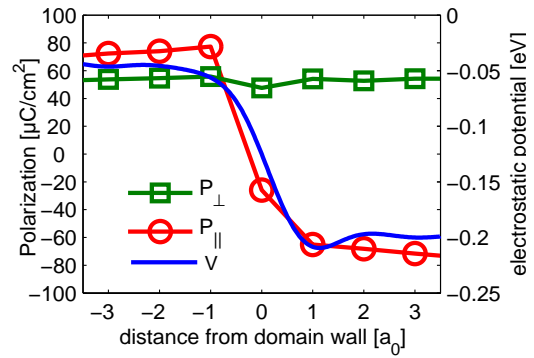}}
\subfigure[]{\includegraphics[scale=0.8]{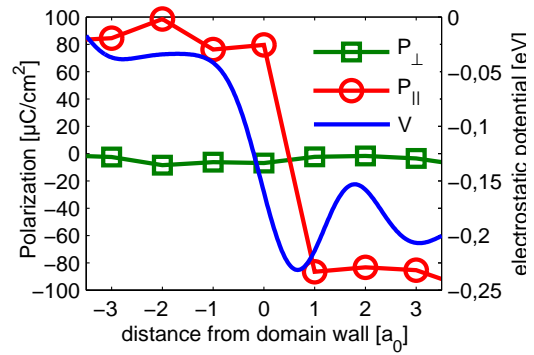}}
\caption{Parallel and normal components of the polarization,
$P_{\parallel}$ and P$_{\perp}$, and the 
macroscopically and planar averaged electrostatic
potential, V for (a) $71^{\circ}$ domain boundary with continuous oxygen octahedral
rotations, (b) $109^{\circ}$ domain boundary with continuous oxygen octahedral
rotations centered on the FeO$_2$ plane and (c) 180\textdegree{} domain boundary with continuous oxygen octahedral
rotations. Note that only half of the supercell is shown. 
\label{fig:pot}}
\end{figure}

Next we analyze the local polarization by decomposing it into the
components parallel and perpendicular to the planes of the domain walls. 
(Figs. \ref{fig:pot}(a), (b) and (c) for 
the 71$^{\circ}$, 109$^{\circ}$ and 180$^{\circ}$ walls respectively.)
The total polarization in the mid-domain regions is $\sim 90\mu$C/cm$^{2}$ 
in all cases, in good agreement with previously reported bulk values 
\cite{Neaton_et_al:2005}. In all cases the component parallel
to the domain wall changes from its full mid-domain value in one
orientation to the full value in the other orientation within two 
or three layers. 
The magnitude of the change in polarization component perpendicular to the 
wall, however, depends strongly on the domain wall type. 
In Figs.~\ref{fig:pot}(a), 
\ref{fig:pot}(b) and \ref{fig:pot}(c) we also plot the planar and 
macroscopically averaged electrostatic potential (extracted as in 
Ref.~\cite{Meyer/Vanderbilt:2002}) across the supercell to illustrate 
the potential step associated with this change in perpendicular 
component of the polarization. For the 
71$^{\circ}$ wall the change in perpendicular component and corresponding
potential step are small; the magnitude of the potential step
is $\sim$0.02 eV. In the 109$^{\circ}$ case the change the in out-of-plane
component is considerable, and the corresponding step is significant
(0.15 eV). This behavior is analogous to that reported previously in
calculations for 90$^{\circ}$ domain walls \cite{Meyer/Vanderbilt:2002}. 
Perhaps surprisingly, the 180$^{\circ}$ boundary shows the largest potential step, of
0.18 eV. (Earlier studies of 180$^{\circ}$ domain boundaries in tetragonal PbTiO$_3$
\cite{Meyer/Vanderbilt:2002} included an inversion center at the domain
wall and therefore obtained no change in perpendicular 
component). The following analysis of the evolution of the polarization 
through successive corners of the pseudocube explains the loss of inversion symmetry and 
the change in the perpendicular component in the 180$^{\circ}$ case.

Interestingly, the presence of the large potential steps at the 109$^{\circ}$ and 
180$^{\circ}$ walls, and the absence of a step at the 71$^{\circ}$ wall, correlate 
with an intriguing recent observation of electrical conductivity at the 109$^{\circ}$
and 180$^{\circ}$ walls, and its absence at the 
71$^{\circ}$ wall \cite{Seidel(2009)}.  A possible explanation of the observed
conductivity is the generation of a space charge layer in the region of the wall to
screen this otherwise energetically unfavorable potential discontinuity. 

\begin{figure}
\subfigure[]{\includegraphics[scale=1.5]{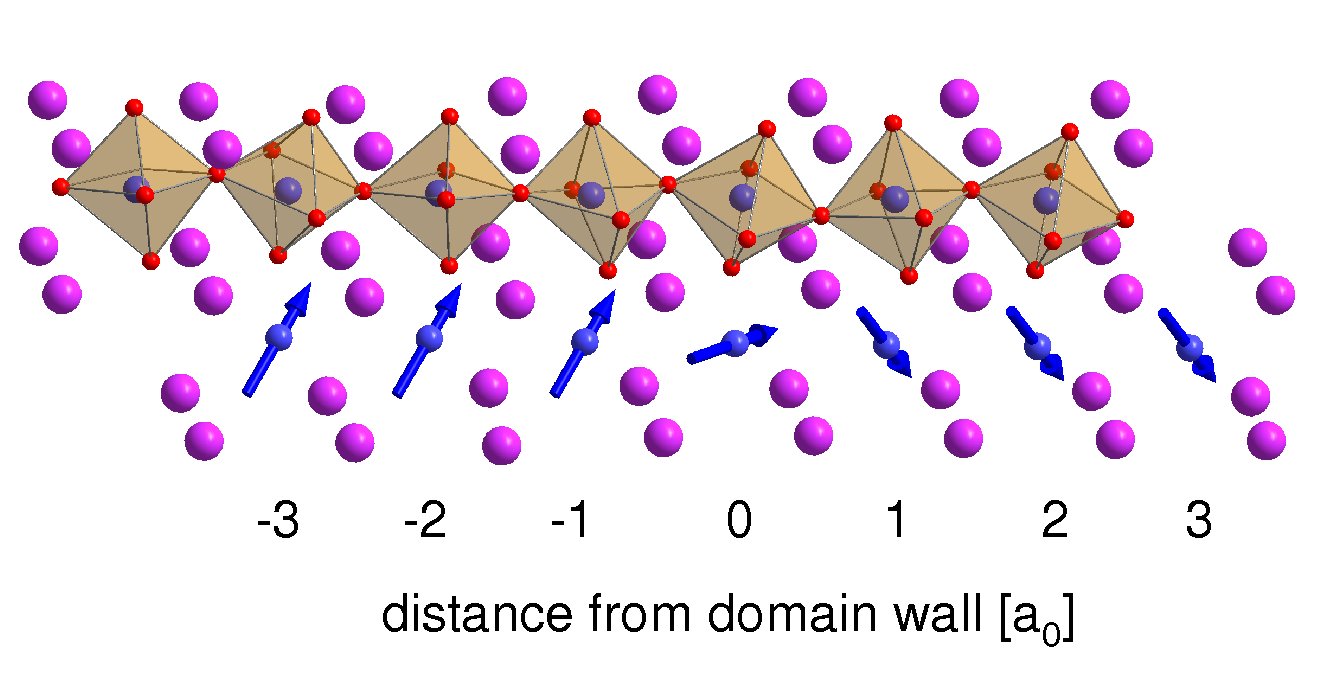}}
\subfigure[]{\includegraphics[scale=1.5]{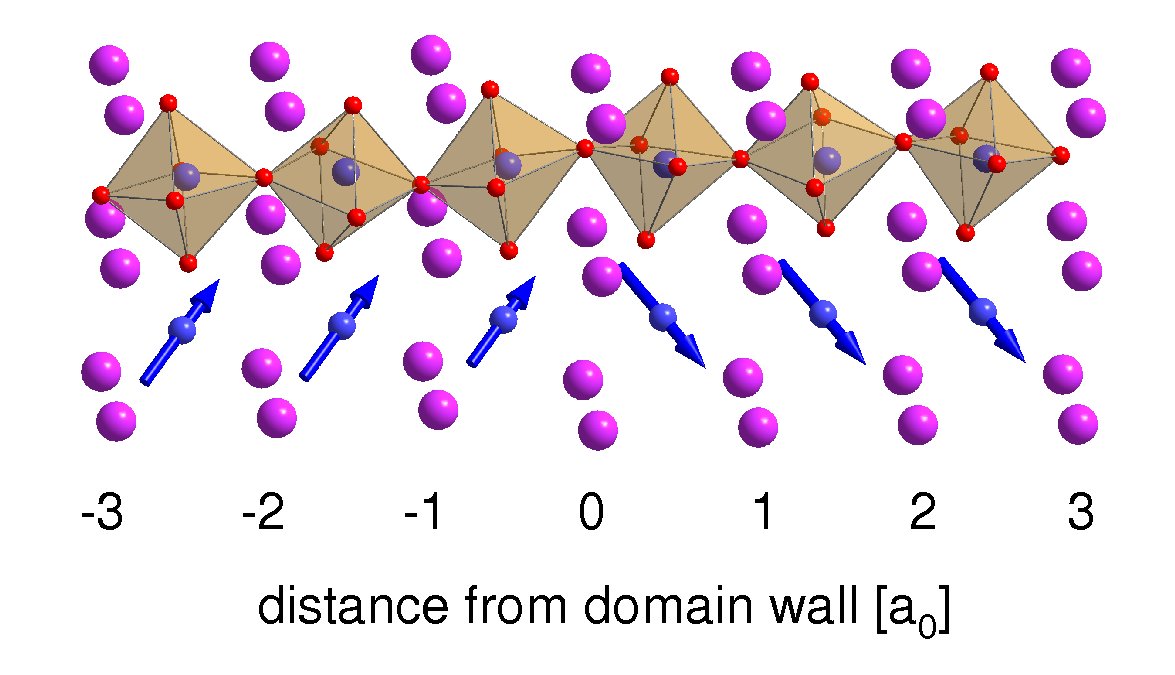}}
\subfigure[]{\includegraphics[scale=1.5]{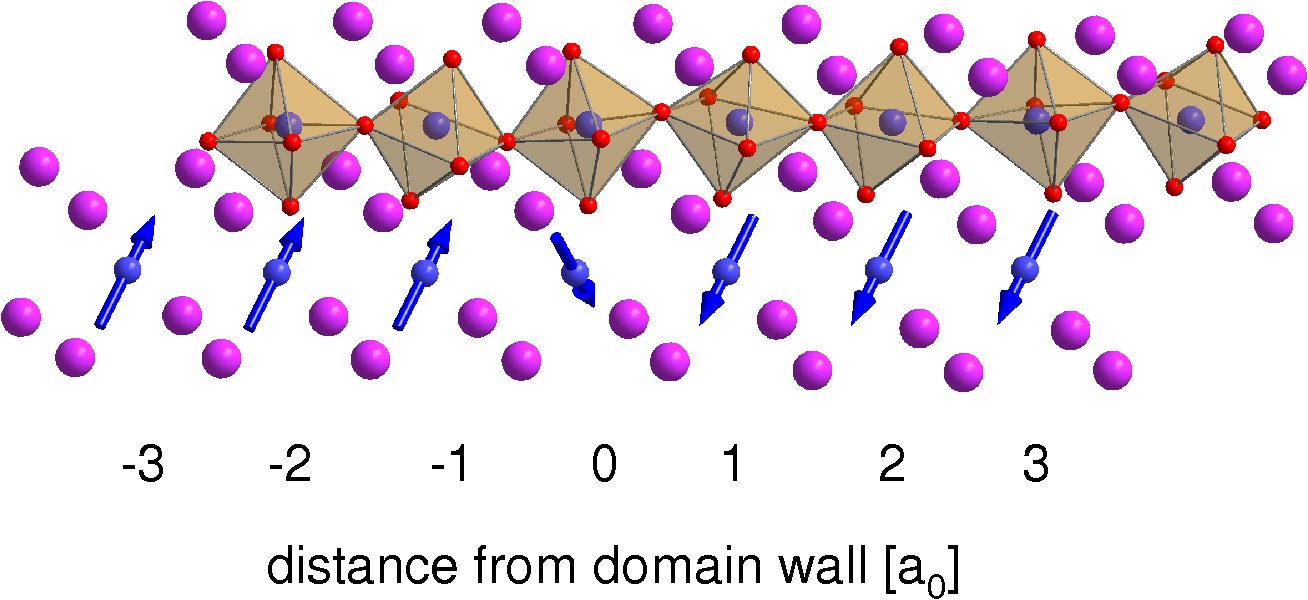}}
\caption{Evolution of the local polarization across (a) the 71\textdegree{}, (b) the  109\textdegree{} and (c) the 180\textdegree{} domain wall.
The blue arrows represent the magnitude and orientation of the local polarization.
\label{fig:dip}}
\end{figure}

Finally, to help with visualizing the change in polarization across the domain walls,
in Fig.~\ref{fig:dip} we indicate the local polarization vectors in each layer of the supercells
as blue arrows showing the magnitude and orientation. In the 71$^{\circ}$ case 
we can clearly see that the polarization rotates from one corner of the
pseudocubic unit cell, through the center of the edge to the adjacent corner,
accompanied by the small attenuation in magnitude  which we saw earlier in
Fig.~\ref{fig:pol}. As already seen in Fig.~\ref{fig:pot}(a), 
this geometry allows the perpendicular component of polarization to remain constant
across the wall. The analogous cartoon for the 180$^{\circ}$ wall 
(Fig.~\ref{fig:dip}(c)) shows that the polarization vector rotates between successive 
corners of the pseudo-cubic unit cell which are the stable orientations of the
polarization in $R3c$ BiFeO$_3$. At the layer-by-layer level of resolution
we see a jump by 71$^{\circ}$ followed by a jump of 109$^{\circ}$; both intermediate
orientations have small components perpendicular to the domain wall. 
Note that imposition of an inversion center during the structural relaxation, 
which might be anticipated for a 180$^{\circ}$ wall, would not have allowed this
ground state to develop. 
In contrast, the change in orientation of the polarization across the 
109$^{\circ}$ wall is accompanied by a rather large attenuation of the total 
polarization (see Fig. \ref{fig:pol} and Fig.~\ref{fig:dip}(b)). 

\subsection{Electronic properties of the domain walls}

In light of the intriguing reported electrical conductivity mentioned above,
we next analyze the electronic properties of the domain walls. We look
particularly at the layer-by-layer densities of states, to see if the structural
deformations in the wall region lead to a closing of the electronic band gap.
Indeed, earlier DFT calculations for bulk BiFeO$_3$ \cite{Neaton_et_al:2005}
indicated a strong dependence of the electronic band gap on the structure. In
particular the ideal cubic structure, in which the 180$^{\circ}$ Fe-O-Fe
bond angles maximize the Fe $3d$ - O $2p$ hybridization and hence the bandwidth,
has a significantly reduced band gap compared with the $R3c$ structure and is 
even metallic within the LSDA.

\begin{figure}
\includegraphics[scale=0.8]{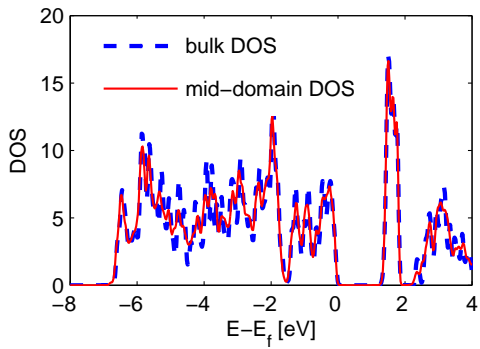}
\caption{Comparison of calculated bulk DOS for BiFeO$_3$ (dashed line) with
the local density of states of a mid-domain layer in the supercell with a 109$^{\circ}$
domain wall. The sum of both spin channels is shown. 
\label{fig:DOSbulk}}
\end{figure}

First, in Figure \ref{fig:DOSbulk} we compare the local density of states (LDOS) 
for a layer in the center of a domain with our calculated density of states for
bulk BiFeO$_3$. The mid-domain LDOS shown is for the supercell containing the 109$^{\circ}$ wall; those of the 71$^{\circ}$ and 180$^{\circ}$ supercells are indistinguishable.
As found in prior work \cite{Neaton_et_al:2005}, the bulk valence band consists of
O $2p$ - majority spin Fe $3d$ hybridized states, while the bottom part of the
conduction band is formed of minority spin Fe $3d$ states and Bi $6p$ states.
The LSDA+$U$ band gap is ~1.4 eV for our chosen values of $U$ and $J$.
The electronic structure in the mid-domain region fully recovers the bulk
behavior.

\begin{figure}
\subfigure[]{\includegraphics[scale=0.8]{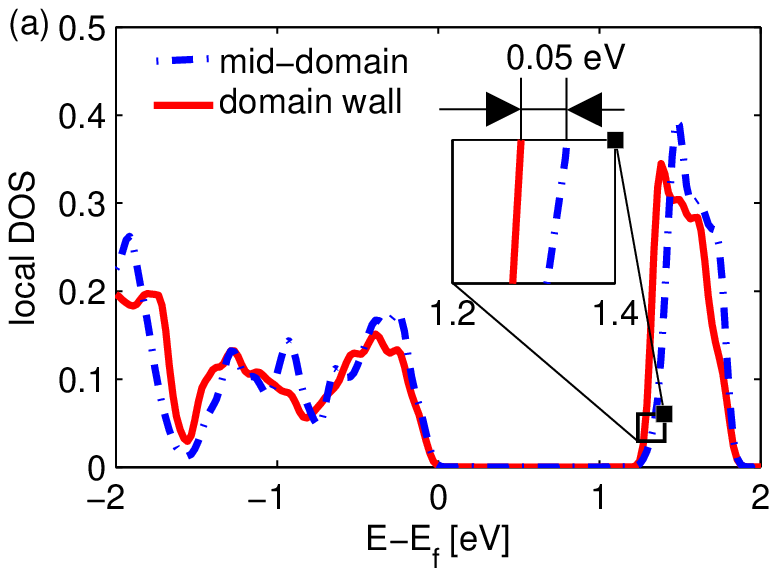}}
\subfigure[]{\includegraphics[scale=0.8]{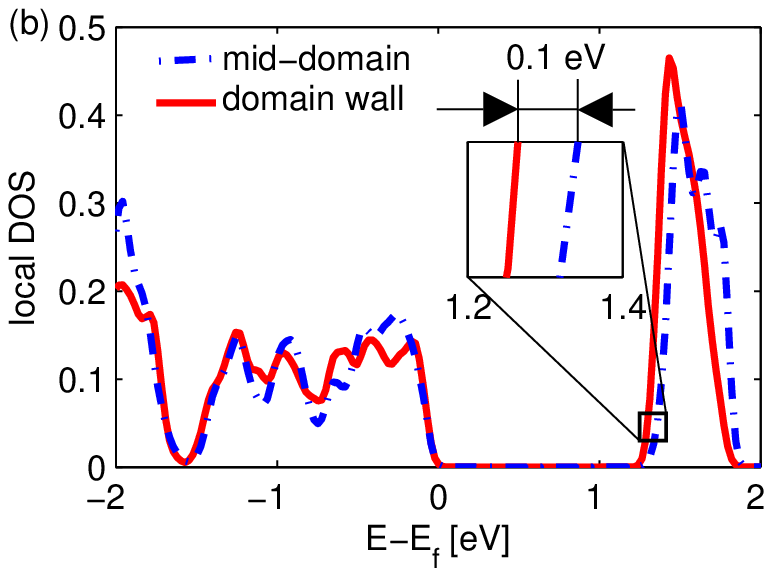}}
\subfigure[]{\includegraphics[scale=0.8]{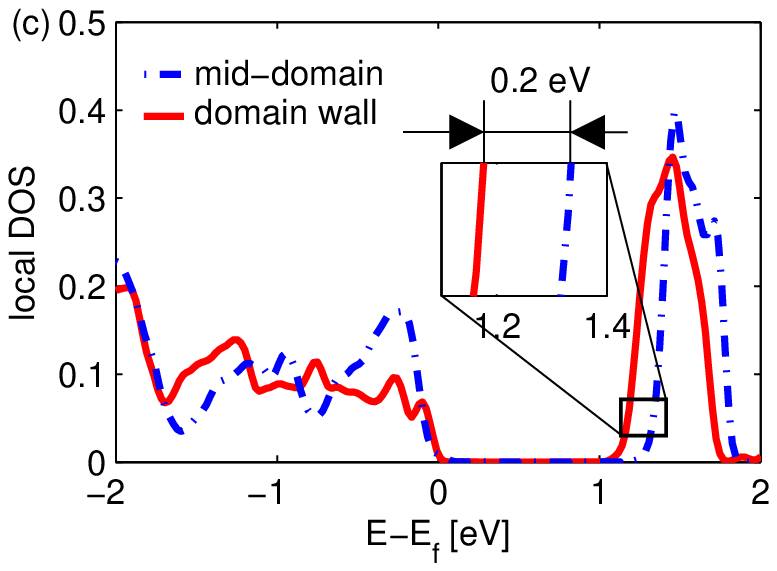}}
\caption{Comparison of the calculated Fe LDOS in the mid-domain and domain
wall regions for (a) 71\textdegree{}, (b) 109\textdegree{} and (c)
180\textdegree{} walls. Note the downward shifts in the conduction band edges, 
particularly in the 109\textdegree{} and 180\textdegree{} cases. 
The 180\textdegree{} case also shows a small upward valence band edge
shift.
\label{fig:partial-Fe-DOS}}
\end{figure}

\begin{figure}
\includegraphics[scale=0.8]{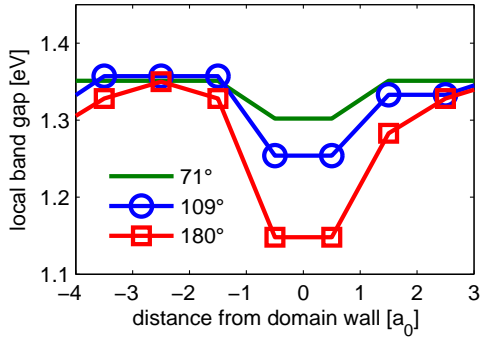}
\caption{Local band gap extracted from the layer-by-layer densities of states.
\label{fig:local-band-gap}}
\end{figure}

In the domain wall, deformation of the Fe-O-Fe angles causes changes in the
hybridization which affect the Fe $e_{g}$ states, resulting in shifts of
the band edges. These are strongest at the 180$^{\circ}$ boundary where the
deformations are largest and the Fe-O-Fe angles are strongly increased. 
Fig.~\ref{fig:partial-Fe-DOS} compares the mid-domain and domain wall 
LDOSs for the three wall orientations. The increasing downward shift in the conduction 
band edge from 71$^{\circ}$ to 109$^{\circ}$ to 180$^{\circ}$ walls,
correlating with the increasing change in Fe-O-Fe band angle is
clearly visible. At the 180\textdegree{} wall 
there is an additional shift of the top of the valence band (consisting of
O $2p$ - Fe $3d$ hybridized states) upwards in energy. 
These band edge shifts in turn cause a reduction in the local band gap,
which is plotted in Fig.~\ref{fig:local-band-gap}.  Again the change is smallest for the 
71\textdegree{} wall and largest for the 180\textdegree{} wall.

In Fig.~\ref{fig:local-band-gap} we show the local band gap extracted from
the layer-by-layer densities of states across the three wall types. In all
cases we see a reduction in the band gap in the wall region, with the 180$^{\circ}$
wall again showing the largest effect. In no case, however, does the gap approach
zero in the wall region. 

\begin{figure}
\subfigure{\includegraphics[scale=0.8]{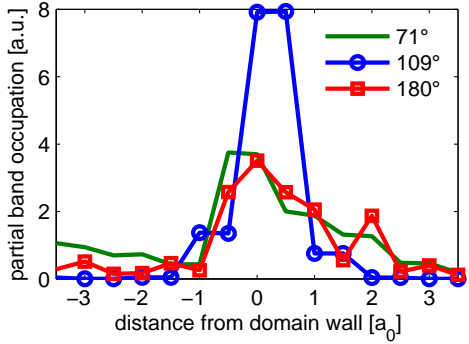}}
\caption{Layer-by-layer projection of the lowest conduction band (partial band 
occupation) for 71\textdegree{}, 109\textdegree{}
and 180\textdegree{} domain walls. The integrated partial band occupation is equal
to one.
\label{fig:lowest-conduction-band}}
\end{figure}

Finally, to provide a quantitative measure of the extent of localization of the states 
near the conduction band edge, we calculate the projection of the lowest energy conduction 
band onto each layer of the supercell; our results are shown in 
Figure~\ref{fig:lowest-conduction-band}. It is clear that the conduction band
edge is dominated by states in the domain wall, more so in the 109\textdegree{}
wall (on the $\left\{ 100\right\}$ plane) than in the 71\textdegree{} and 180\textdegree{}
walls (which are both in $\left\{ 110\right\}$ planes). Again, this implies
that electron carriers in the system, which will occupy the lowest conduction
band states, will accumulate at the domain boundary regions.

\subsection{Magnetic properties}
\label{Magnetism}

In bulk BiFeO$_3$, the magnetic ordering is G-type antiferromagnetic
with a long wavelength ($\sim620$ \AA) spiral of the AFM axis 
\cite{Sosnowska/Peterlin-Neumaier/Streichele:1982}. The spiral is
known to be suppressed by doping \cite{Sosnowska_et_al:2002} and 
importantly for this work is believed to be suppressed in thin films
\cite{Bai_et_al:2005}. Our earlier first-principles calculations
showed that the AFM vector lies in one of six easy axes within 
the magnetic easy plane which is perpendicular to the polar axis.
We found a spin-orbit driven canting of the magnetic moments of $\sim$1$^{\circ}$
\cite{Ederer/Spaldin:2005} which, in the absence of a spiral, results in
a net weak ferromagnetism of 0.05 $\mu_{\rm B}$ per Fe ion. The canting is 
symmetry allowed because of the
octahedral rotations; a hypothetical $R3m$ polar structure without octahedral
rotations could not show weak ferromagnetism.  Recent magneto-optical measurements 
showed the antiferromagnetism can be controlled using an electric field because 
its orientation is  determined by the direction of the ferroelectric 
polarization \cite{Zhao_et_al:2006}.

In this final section we include spin-orbit coupling in our calculations
in order to explicitly calculate the orientation of the magnetic moments
relative to the polarization vector, and to allow any spin-orbit driven
canting to manifest. Consistent with Ref.~\cite{Bai_et_al:2005} and for
computational feasibility, we use the ideal G-type structure with initial spin polarization
axis set to the pseudocubic $[1\bar{1}0]$ direction, which was chosen because
it is perpendicular to the electric polarization vectors on both sides of the
domain wall ($[111]\,/\,[\bar{1}\bar{1}1]$ in the $109^{\circ}$ case) as our starting
point; we do not allow the long wavelength spiral.
Since the non-collinear calculations with spin-orbit coupling are so
computationally intensive, we are only able to study one wall orientation.
We choose to study the 109$^{\circ}$ wall
since it is accompanied by  a reorientation of the antiferromagnetic 
easy plane across the boundary; our findings might also be applicable 
to the 71$^{\circ}$ domain wall in which the easy plane also reorients
across the boundary.
(At the 180$^{\circ}$ domain 
wall the polarization reverses direction and so we expect the easy plane 
of magnetization, which is perpendicular to the polarization, to remain 
unchanged across the domain wall. In addition, since our structural studies
described above found that the phase of the octahedral rotations -- which
determines the orientation of the canting -- is unchanged across the domain
wall, we do not expect a reversal of the weak ferromagnetic vector.)
We expect that changes in the local symmetry at the wall might significantly 
affect the canting angles; in addition if the perturbations in the Fe-O-Fe 
bond angles are large enough we could even see a change from antiferromagnetic to
ferromagnetic superexchange 
\cite{Kanamori:1959,Anderson:1963,Goodenough:book}.

\begin{figure}
\subfigure{\includegraphics[scale=1.5]{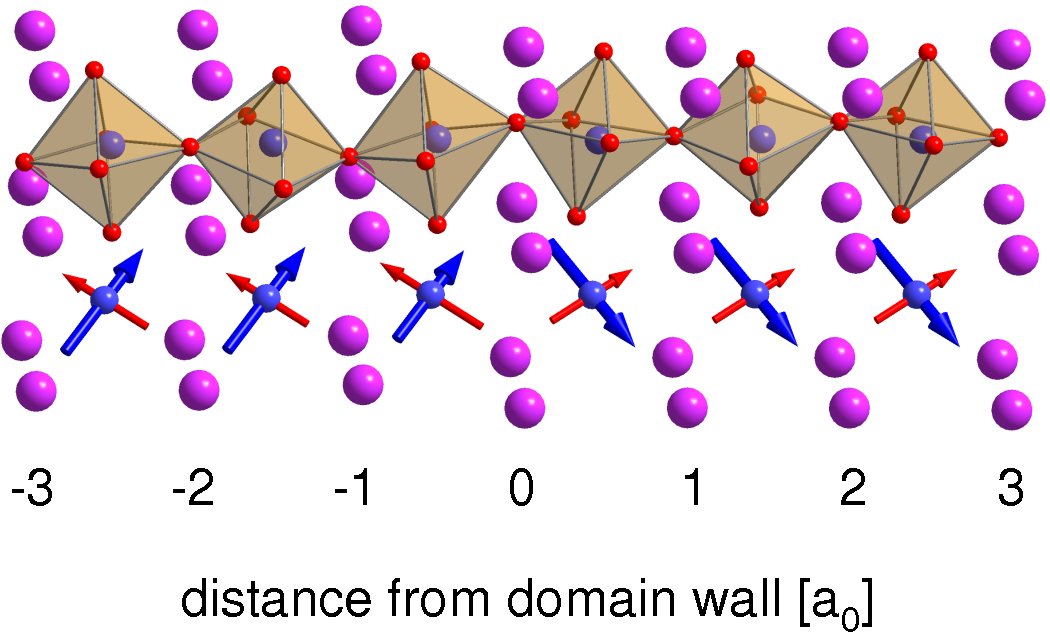}}
\subfigure{\includegraphics[scale=0.8]{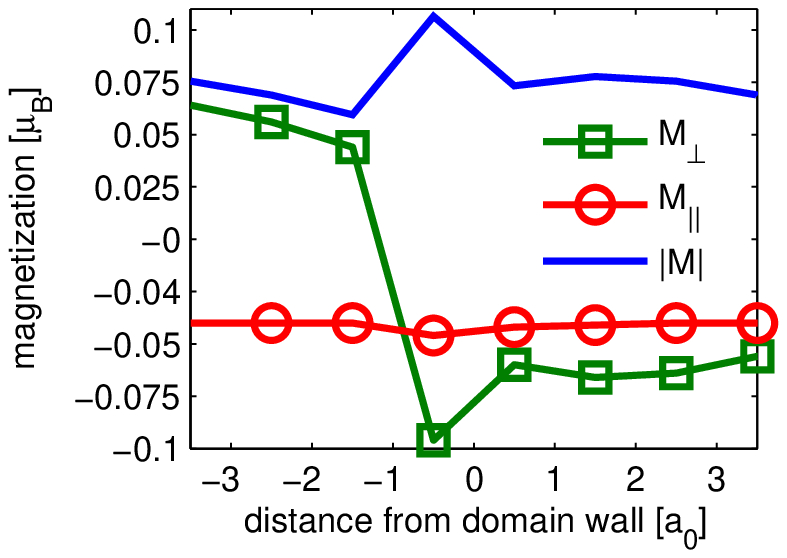}}
\caption{Layer-by-layer local magnetic moment across the 109\textdegree{} 
domain boundary. (a) Shows the local magnetization vectors (small red arrows) resulting from
the canting of the Fe magnetic moments in each layer, which is perpendicular to the 
local electric polarization (big blue arrow),
(b) shows the local component of the magnetization projected parallel and perpendicular 
to the wall plane, and the local magnitude.
\label{fig:net-spin-across-109=00003D0000B0}}
\end{figure}

In Fig.~\ref{fig:net-spin-across-109=00003D0000B0} we show the net
local magnetization resulting from the canting of the Fe magnetic moments
in each layer across the 109$^{\circ}$ domain wall. In the mid-domain regions
the orientation of the local moment is $[11\bar{2}]$ on one side and $[112]$ on 
the other side of the wall. 
The magnitude of the local moment is consistent with that calculated for bulk
BiFeO$_3$. The reorientation of the AFM plane consistent with the reorientation
of the polarization is evident. Importantly, we see that the local
canting {\it increases} by $\sim$33\% in the wall layer, consistent with the larger
deviation of the Fe-O-Fe angles from 180$^{\circ}$.
This behavior could explain the intriguing recent observation that the magnitude
of the exchange bias in BiFeO$_3$/Co multilayers is affected by the ferroelectric
domain structure in BiFeO$_3$ \cite{Martin_et_al:2009}.

\section{Summary\label{Summary}}
In summary, we have used the LSDA$+U$ method of density functional theory
to calculate the structural,
electronic and magnetic properties of the ferroelectric domain walls in
multiferroic BiFeO$_3$. We have identified the wall orientations that
are most likely to occur based on their relative energy costs; in particular
we have shown that walls in which the rotations of 
the oxygen octahedra do not change their phase when the polarization 
reorients are significantly more favorable than those with rotation
discontinuities. Our analysis of the local polarization and electronic
properties revealed potential steps and reduction in local band gaps
at the 109$^{\circ}$ and 180$^{\circ}$ walls; these correlated with
recent measurements of electrical conductivity at these boundaries.
Finally, we showed that changes in structure at the domain walls
cause changes in canting of the Fe magnetic moments which
can enhance the local magnetization at the domain walls. The latter
suggests possible new routes to electric field-control of magnetism
in BiFeO$_3$.

\section{Acknowledgments}

Spaldin was supported by the National Science Foundation under Award No. DMR-0605852.
Calculations were performed at the San Diego Supercomputer Center, and at the National 
Center for Supercomputer Applications. We furthermore acknowledge the DFG for funding through FOR 520 and Ge 1202/5-1 and the BMBF for funding via the Pakt fuer Forschung und Innovation.

\bibliography{manuscript}

\end{document}